\documentclass[twoside,twocolumn,9pt]{article}
\usepackage{extsizes}
\usepackage[super,sort&compress,comma]{natbib} 
\usepackage[version=3]{mhchem}
\usepackage[left=1.5cm, right=1.5cm, top=1.785cm, bottom=2.0cm]{geometry}
\usepackage{balance}
\usepackage{mathptmx}
\usepackage{sectsty}
\usepackage{graphicx} 
\usepackage{lastpage}
\usepackage[format=plain,justification=justified,singlelinecheck=false,font={stretch=1.125,small,sf},labelfont=bf,labelsep=space]{caption}
\usepackage{float}
\usepackage{fancyhdr}
\usepackage{fnpos}
\usepackage[english]{babel}
\addto{\captionsenglish}{%
  
}
\usepackage{array}
\usepackage{droidsans}
\usepackage{charter}
\usepackage[T1]{fontenc}
\usepackage[usenames,dvipsnames]{xcolor}
\usepackage{setspace}
\usepackage[compact]{titlesec}
\usepackage{hyperref}

\usepackage{epstopdf}
\usepackage{float}

\newcommand{\be}{\begin{equation}}

\newcommand{\We}{\rm{We}}
\newcommand{\Oh}{\rm{Oh}}

\definecolor{cream}{RGB}{222,217,201}

\begin{document}

\pagestyle{fancy}
\thispagestyle{plain}
\fancypagestyle{plain}{
\renewcommand{\headrulewidth}{0pt}
}

\makeFNbottom
\makeatletter
\renewcommand\LARGE{\@setfontsize\LARGE{15pt}{17}}
\renewcommand\Large{\@setfontsize\Large{12pt}{14}}
\renewcommand\large{\@setfontsize\large{10pt}{12}}
\renewcommand\footnotesize{\@setfontsize\footnotesize{7pt}{10}}
\makeatother

\renewcommand{\thefootnote}{\fnsymbol{footnote}}
\renewcommand\footnoterule{\vspace*{1pt}%
\color{cream}\hrule width 3.5in height 0.4pt \color{black}\vspace*{5pt}} 
\setcounter{secnumdepth}{5}

\makeatletter 
\renewcommand\@biblabel[1]{#1}            
\renewcommand\@makefntext[1]%
{\noindent\makebox[0pt][r]{\@thefnmark\,}#1}
\makeatother 
\renewcommand{\figurename}{\small{Fig.}~}
\sectionfont{\sffamily\Large}
\subsectionfont{\normalsize}
\subsubsectionfont{\bf}
\setstretch{1.125} 
\setlength{\skip\footins}{0.8cm}
\setlength{\footnotesep}{0.25cm}
\setlength{\jot}{10pt}
\titlespacing*{\section}{0pt}{4pt}{4pt}
\titlespacing*{\subsection}{0pt}{15pt}{1pt}

\fancyfoot{}
\fancyfoot[LO,RE]{\vspace{-7.1pt}\includegraphics[height=9pt]{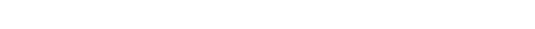}}
\fancyfoot[CO]{\vspace{-7.1pt}\hspace{13.2cm}\includegraphics{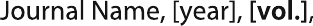}}
\fancyfoot[CE]{\vspace{-7.2pt}\hspace{-14.2cm}\includegraphics{head_foot/RF}}
\fancyfoot[RO]{\footnotesize{\sffamily{1--\pageref{LastPage} ~\textbar  \hspace{2pt}\thepage}}}
\fancyfoot[LE]{\footnotesize{\sffamily{\thepage~\textbar\hspace{3.45cm} 1--\pageref{LastPage}}}}
\fancyhead{}
\renewcommand{\headrulewidth}{0pt} 
\renewcommand{\footrulewidth}{0pt}
\setlength{\arrayrulewidth}{1pt}
\setlength{\columnsep}{6.5mm}
\setlength\bibsep{1pt}

\makeatletter 
\newlength{\figrulesep} 
\setlength{\figrulesep}{0.5\textfloatsep} 

\newcommand{\topfigrule}{\vspace*{-1pt}%
\noindent{\color{cream}\rule[-\figrulesep]{\columnwidth}{1.5pt}} }

\newcommand{\botfigrule}{\vspace*{-2pt}%
\noindent{\color{cream}\rule[\figrulesep]{\columnwidth}{1.5pt}} }

\newcommand{\dblfigrule}{\vspace*{-1pt}%
\noindent{\color{cream}\rule[-\figrulesep]{\textwidth}{1.5pt}} }

\makeatother

\twocolumn[
  \begin{@twocolumnfalse}
{\includegraphics[height=30pt]{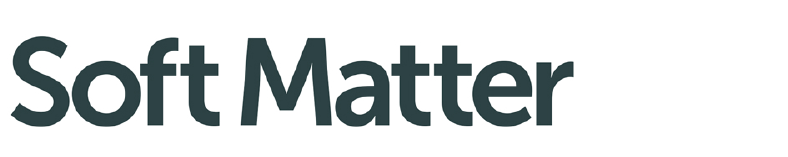}\hfill\raisebox{0pt}[0pt][0pt]{\includegraphics[height=55pt]{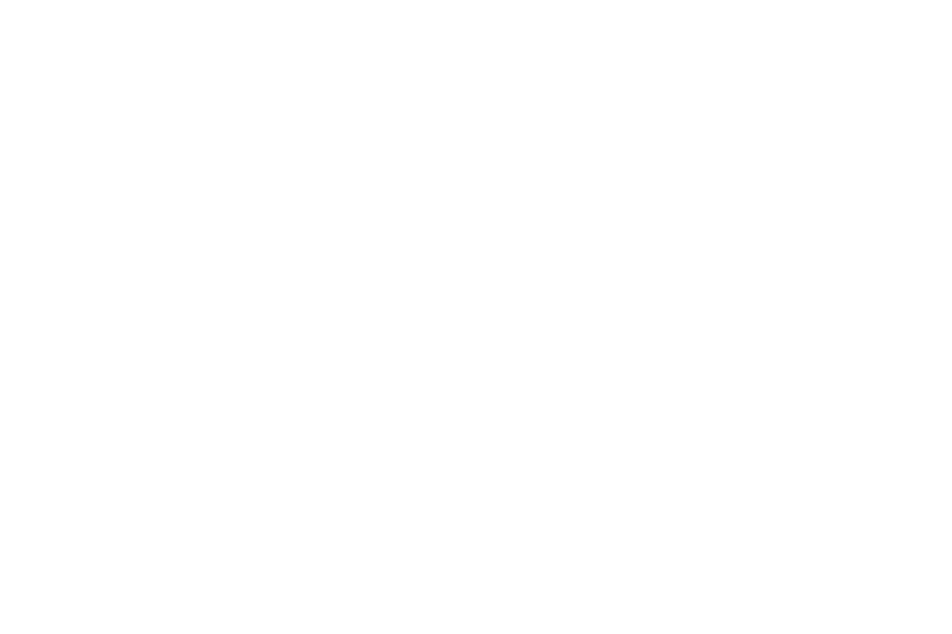}}\\[1ex]
\includegraphics[width=18.5cm]{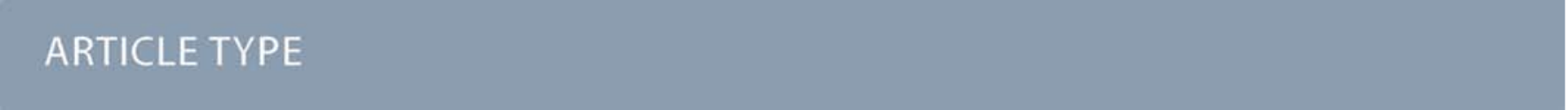}}\par
\vspace{1em}
\sffamily
\begin{tabular}{m{4.5cm} p{13.5cm} }

\includegraphics{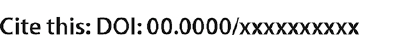} & \noindent\LARGE{\textbf{Viscous droplet impingement on soft substrates$^\dag$}} \\
\vspace{0.3cm} & \vspace{0.3cm} \\

 & \noindent\large{Marcus Lin,\textit{$^{a}$} Quoc Vo,\textit{$^{a}$} Surjyasish Mitra,\textit{$^{b}$} and Tuan Tran$^{\ast}$\textit{$^{a,b}$}} \\

\includegraphics{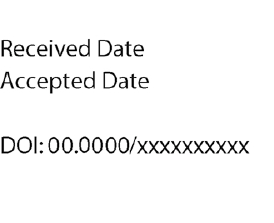} &\noindent\normalsize{
Viscous droplets impinging on soft substrates may exhibit several distinct behaviours
including repeated bouncing, wetting, and hovering, i.e.,  
spreading and retracting after impact without bouncing back or wetting. 
We experimentally study the conditions enabling these characteristic 
behaviours by systematically varying the substrate elasticity,
impact velocity
and the liquid viscosity.
For each substrate elasticity, the transition to wetting 
is determined 
as the dependence of the Weber number $\We$, 
which measures the droplet's
kinetic energy against its surface energy,
on the Ohnesorge number $\Oh$, which 
compares viscosity to inertia and capillarity.
We find that while $\We$ at the wetting 
transition monotonically decreases with 
$\Oh$ for relatively rigid substrates, 
it exhibits
a counter-intuitive behaviour in which 
it first increases 
then gradually decreases for softer substrates. 
We experimentally determine the 
dependence of the maximum Weber number 
allowing non-wetting impacts on the substrate elasticity and show 
that it
provides an excellent quantitative measure 
of liquid repellency for a wide range of surfaces, 
from liquid to 
soft surfaces and non-deformable surfaces. 
}\\
\end{tabular}

 \end{@twocolumnfalse} \vspace{0.6cm}

  ]

\renewcommand*\rmdefault{bch}\normalfont\upshape
\rmfamily
\section*{}
\vspace{-1cm}


\footnotetext{\textit{$^{a}$School of Mechanical and Aerospace Engineering, Nanyang Technological University, Singapore}}
\footnotetext{\textit{$^{b}$Division of Physics and Applied Physics, School of Physical and Mathematical Sciences, Nanyang Technological University, Singapore}}




\section{Introduction}
Droplets impingement on solid surfaces occurs frequently, 
both in nature, e.g., 
raindrops falling on leaves,
and technological applications, 
e.g.,
spray coating   
and inkjet printing \cite{barthlott_purity_1997,de_gans_inkjet_2004,jia_experimental_2003,van_der_bos_velocity_2014}. 
Resulting phenomena --- 
splashing, wetting, bouncing and hovering --- 
occur largely depending 
on the impact velocity: 
while wetting and splashing happen for droplets impacting at high velocity \cite{josserand_drop_2016,howland_its_2016},
bouncing and hovering  
are often observed for those at low velocity 
\cite{chen_droplet_2016,ajaev_levitation_2021,mitra_bouncing--wetting_2021}.
Detailed and mechanistic studies of these phenomena 
reveal that they are closely related to stability of 
the air film separating an impacting droplet and a solid surface. 
Wetting is initiated when the air film 
\cite{thoroddsen_air_2005,bouwhuis_maximal_2012,langley_droplet_2020}
is ruptured due to relatively high impact velocity. 
At higher impact velocity, 
the lubrication pressure built-up in the air film
is sufficiently high to cause splashing, or ejection of spreading lamella
\cite{xu_drop_2005,mandre_precursors_2009,driscoll_ultrafast_2011}.
In contrast, 
an impacting droplet rebounds 
or hovers above the surface  
for low velocity impacts. 
In this case, the air film is stable and sustained
for a significant duration, allowing 
the droplet to exhibit non-wetting impact behaviours,
similar to water droplets on superhydrophobic surfaces
\cite{chen_bouncing_2010,kolinski_skating_2012,van_der_veen_direct_2012,de_ruiter_dynamics_2012,de_ruiter_wettability-independent_2015,ruiter_bouncing_2015}.
The transition between the two behavioural extremes, non-wetting 
and wetting droplet impacts, is crucial for either
fabrication of surfaces having special functionalities, 
e.g., self-cleaning,
contamination resistant \cite{jung_dynamic_2008,bhushan_natural_2011}, 
or applications involving high-resolution 3D printing of biomaterials 
\cite{li_review_2018}.
In most of these applications, 
there are two notable deviations from the bulk of droplet impact studies: 
the impacted surfaces 
are not rigid but deformable, and the viscosity of the working liquid
is significantly high. 
The expansion of the parameter space 
to include 
surface's deformity and 
high liquid viscosity generates
numerous 
impact outcomes, many of which remain unexplored. 

\begin{figure*}[h]
    \centering
    \includegraphics[width=2\columnwidth]{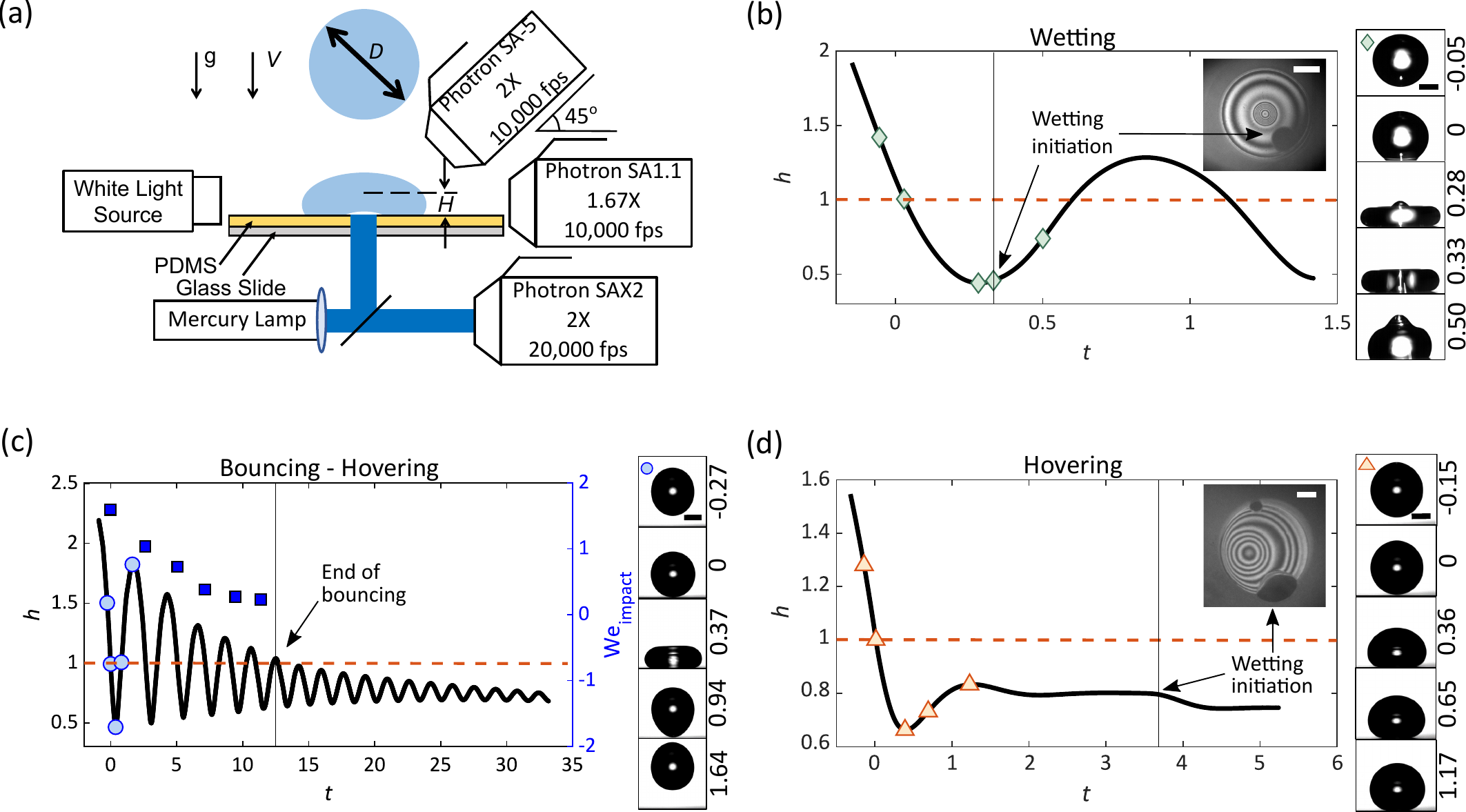}
    \caption{\label{figure1}(a) Schematic of the experimental setup. 
    Droplets of diameter $D$ impacting a soft substrate 
    with velocity $V$. The impact behaviour is 
    recorded synchronously 
    using top-view and side-view cameras, while
    wetting initiation is detected using an
    interferometric bottom-view camera. 
    (b) Normalised centre-of-mass height $h$
    vs. dimensionless time $t$ 
    for an impacting droplet in the wetting regime. 
    The impact conditions are $E = 30\,$kPa,
    $\Oh = 0.011$ and $\We = 9.8$. 
    The vertical panel shows side-view snapshots of the droplet during impact. 
    (c) $h$ vs. $t$ for an impacting droplet in the bouncing-hovering regime
    for $E = 30\,$kPa,
    $\Oh = 0.011$ and $\We = 1.6$. 
    The right axis indicates 
    the Weber number at impact 
    of droplet in the bouncing behaviour (blue squares).
    (d) $h$ vs. $t$ for an impacting droplet in the hovering regime
    for $E = 30\,$kPa,
    $\Oh = 0.745$ and $\We = 1.4$. 
    The interferometric snapshots
    in the insets of (b) and (d) are taken  
    when the air film separating the droplet and surface ruptures.
    All scale bars represent 1$\,$ mm.}
\end{figure*}

On one hand, a solid substrate
deforms under impact of a droplet
when 
its rigidity
is reasonably low. 
Deformation of soft substrates with Young's modulus 
in the range $5\,\rm{kPa} - 500\,\rm{kPa}$
was found capable of 
suppressing splashing
for high-velocity droplet impacts \cite{howland_its_2016}.
This is due to early-time substrate deformation,
which inhibits break-up of the spreading lamella
and prevents it from getting 
ejected from the surface. 
At low impact velocity, various impact phenomena 
were studied 
for water droplets 
impacting substrates with 
shear modulus in the range 
 $0.2\,\rm{kPa} - 510\,\rm{kPa}$
\cite{chen_droplet_2016}. 
For surfaces at the lower range of shear modulus, 
the observed velocity at the transition to wetting increases with the shear modulus
for low viscosity liquids,
implying that it is more difficult to induce wetting on softer surfaces. 
In addition, the range of impact velocity for droplet rebound
is wider for soft surfaces compared to their rigid counterparts. 
Interestingly, for water droplets impacting on soft surfaces, 
air cavities formed in the liquid bulk due to capillary-induced waves 
were observed near the transition to wetting, 
suggesting that air cavity formation 
may alter the wetting mechanism for droplets of low viscosity liquids
\cite{chen_droplet_2016,mitra_bouncing--wetting_2021}.

On the other hand, 
the dependence of 
the transition to wetting 
on liquid viscosity of impacting droplets 
has not been conclusive, even for the case of impacts 
on rigid surfaces. 
First, for droplets having viscosity $\mu\le109\,$mPa$\,$s impacting on glass surfaces,
the transition to wetting was shown to weakly depend on 
the viscosity \cite{de_ruiter_wettability-independent_2015}.
Nonetheless,
a dependence on viscosity for transition to wetting 
at low impact velocity is observed
when the viscosity range is expanded to 200$\,$mPa$\,$s \cite{jha_viscous_2020}.
When droplet viscosity is approximately 200$\,$mPa$\,$s, 
viscous effect becomes dominant and inhibiting to bouncing \cite{jha_viscous_2020}.
On soft substrates, the transition to wetting for impacting droplets of
high viscosity liquids has not been investigated. 

In this paper, we experimentally study the behaviours 
of viscous droplets impinging on soft substrates
by systematically varying the substrate elasticity $E$, 
impact velocity $V$ and liquid viscosity $\mu$. 
We then investigate how $\mu$ 
influences the transition to wetting 
while varying the substrate elasticity $E$.
We explore the connection between air cavity formation in the bulk of impacting droplets 
and the transition to wetting, in particular how this connection changes 
for impacts on softer substrates. 
Finally, we quantify the ability of soft substrates 
in repelling droplets 
by determining the maximum Weber number allowing non-wetting behaviour 
in the explored ranges of substrate elasticity and liquid viscosity.

\section{Methods and Materials}
To fabricate soft 
substrates on glass slides, we use Polydimethylsiloxane (PDMS) 
(Dow Corning, Sylgard$\textsuperscript{\textregistered}$184).
The monomer to cross-linker weight ratio 
 is varied between 50:1 and 10:1;
correspondingly, the resulting Young's modulus $E$ 
of the substrate is varied from 
30$\,$kPa to 900$\,$kPa.
For each batch of fabricated substrates, 
we measure the Young's modulus 
of a representative sample 
by a rheometer (Discovery Hybrid, TA Instrument).
The thickness of all substrates is kept fixed at $\approx2\,$mm 
to ensure negligible end effects caused by the glass slides \cite{style_universal_2013}.

Fig.\:\ref{figure1}a shows 
the schematic of our experimental setup. 
We generate droplets by using a syringe pump
to dispense 
glycerol-water mixture from a flat-tipped needle.
We set the pumping rate at 1$\,\mu$l/min,  
sufficiently small to cause droplet detachment 
from the needle by gravitational force.
The droplet diameter $D$ is kept fixed at 
2.66 $\pm$ 0.14$\,$mm. 
The viscosity $\mu$ of glycerol-water mixtures 
is varied between 5.0$\,$mPa$\,$s to 905.7$\,$mPa$\,$s 
\cite{cheng_formula_2008,volk_density_2018}.
As a result, the corresponding Ohnesorge number, 
defined as $\Oh =\mu(\rho\gamma\textit{D})^{-1/2}$,
varies between 0.011 and 1.973. 
Here, $\rho$ and $\gamma$ respectively are the density 
and surface tension of the mixtures.
Once a droplet is detached from the needle, it impinges
on a soft substrate with impact velocity $V$, which is varied 
from 0.17$\,$ms$^{-1}$ to 0.60$\,$ms$^{-1}$
by adjusting the needle's height.
Correspondingly, the Weber number,
defined as $\We = \rho \textit{V}^2 \textit{D}/\gamma$, 
changes from 0.9 to 16.0.

We record impact behaviours of all droplets 
from top, bottom and side views 
using three synchronous 
high-speed cameras (SA-5, SA-X2 and SA1.1, Photron) 
operating at 10,000, 10,000 and 20,000 frames per second, respectively (see Fig.~\ref{figure1}a).
The top-view camera,
used to observe air cavity formation in the liquid bulk, 
is positioned to record impacting droplets 
at an angle of 45$^\circ$ to the horizontal direction. 
A $2\times$ objective was used with the top-view camera 
to provide a spatial resolution of 10$\,\mu$m per pixel.
The side-view camera 
is connected to 
a $1.67\times$ objective to obtain a spatial resolution of 11.97$\,\mu$m per pixel. 
The bottom-view camera, illuminated coaxially 
using a mercury lamp with a mercury line bandpass filter (wavelength $\lambda = 436\,$nm), 
is used to obtain interferometric recordings of the air films between 
impacting 
droplets and soft substrates.
A $2\times$ objective was used with the bottom-view camera 
to provide a spatial resolution of 10$\,\mu$m per pixel.
We use bottom-view recordings to detect wetting initiation 
during impacts of droplets (insets in Fig.~\ref{figure1}b and d).

\section{Results and Discussions}
In this section, we first examine 
characteristic impact behaviours
observed 
with varying impact velocity $V$, droplet viscosity $\mu$, 
and substrate elasticity $E$. 
We then focus on the effect of substrate elasticity on 
the transition between the two major behaviours:
wetting and non-wetting.
Finally, we discuss the maximum 
Weber number allowing non-wetting behaviour
and how it depends on the substrate elasticity. 

\subsection{Characteristic impact behaviours of droplets on soft substrates}
We observe three characteristic behaviours during impacts
of droplets on soft substrates. 
In the \emph{wetting} behaviour, 
a droplet impinging on a soft substrate 
first reaches its maximum deformation 
before wetting initiates. 
In Fig.\:\ref{figure1}b, 
we show a representative plot of 
the dimensionless 
centre-of-mass height, defined as $h = 2H/D$, 
versus the dimensionless time $t\,=\,T/\tau$. 
Here, the centre-of-mass height $H$ of a droplet 
during impact is determined using 
side-view recordings, assuming that the droplet is axisymmetrical; 
$\tau = \zeta (\rho D^3/8\gamma)^{1/2}$ is the Rayleigh timescale 
with the prefactor $\zeta=2.6\pm 0.1$ \cite{richard_contact_2002}.
The specific conditions in this experiment are 
$\Oh = 0.011$, $\We = 9.8$ and  
$E = 30\,\rm{kPa}$.
We note that
wetting initiation at $t = 0.33$ is detected using 
bottom-view interferometry. 

In the so-called \emph{bouncing-hovering} behaviour, 
a droplet 
impacts and bounces multiple times 
before transitioning into the so-called hovering state in which 
it periodically deforms without 
detaching from the substrate.
We note that the droplet transitions 
from bouncing to hovering when the 
peak value $h_p$ within one impact cycle of $h$
becomes smaller than 1,
as exemplified in Fig.\:\ref{figure1}c.
In this particular example,
both $\Oh$ and $E$ are the same as those shown in 
Fig.\:\ref{figure1}b, but the Weber number is 
reduced from $\We = 9.8$ to $1.6$. 
The droplet bounces several times with decreasing $h_p$ 
due to viscous dissipation within the droplet.
Eventually $h_p$ becomes smaller than 1, 
signifying substantial loss of total kinetic energy to viscous dissipation. 
It is remarkable that although the bottom surface of the droplet 
does not appear to leave the soft surface during hovering, 
its periodic deformation 
enables the air film separating 
the droplet and the soft surface to replenish and sustain 
for a significant duration.
For the experiment shown in 
Fig.\:\ref{figure1}c, the time to wetting is $t\approx50$, considerably larger than that
in the wetting behaviour ($t=0.33$). 

\begin{figure}[t!]
    \centering
    \includegraphics[width=\columnwidth]{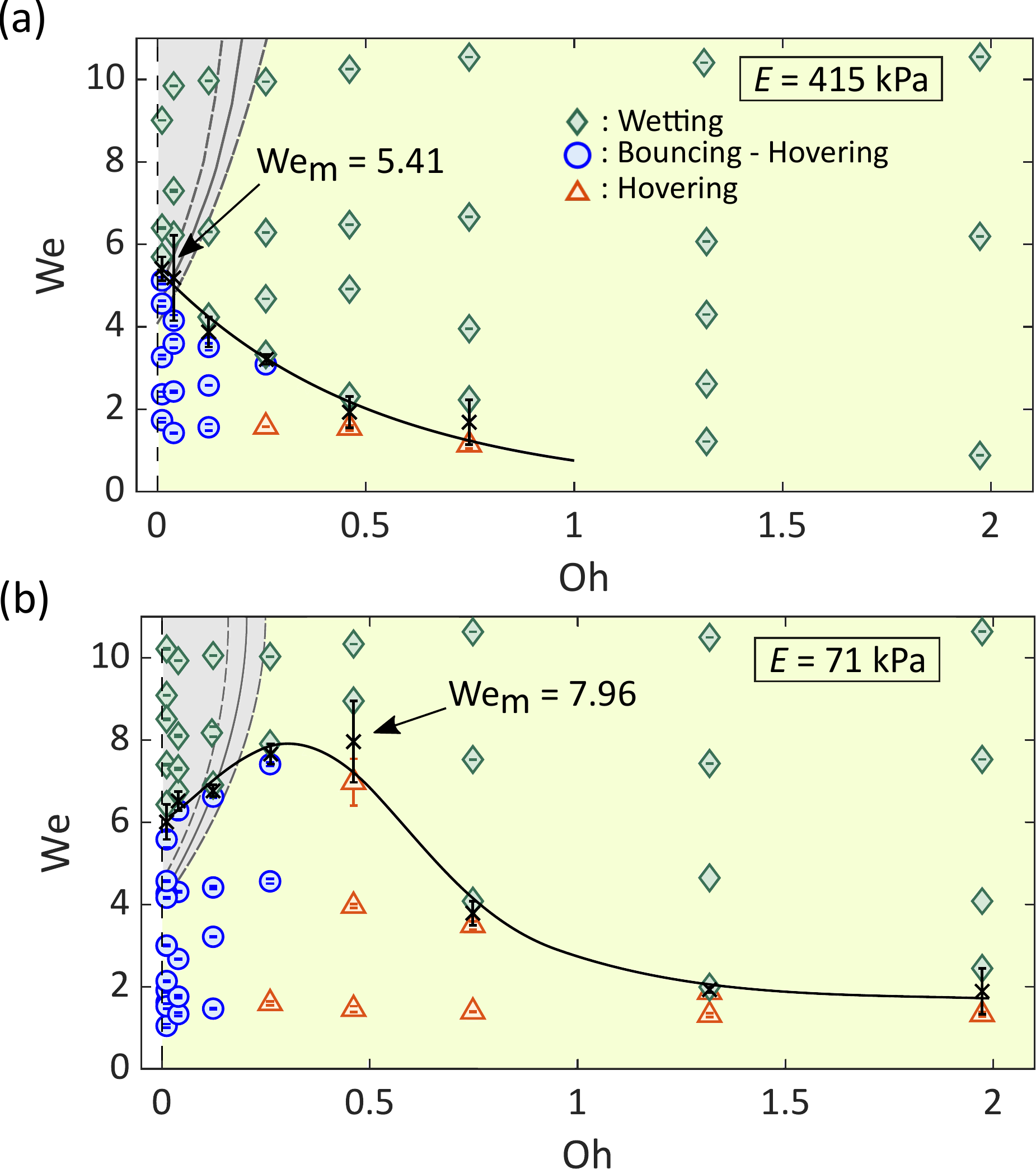}
    \caption{\label{figure2}
    Phase diagrams of impact behaviours
    for glycerol-water droplets 
    falling on soft substrates with (a) $E = 415\,k$Pa and (b) $E = 71\,$kPa. 
    Here, the non-wetting behaviours, 
    which include bouncing-hovering (blue circles)
    and hovering (orange triangles), 
    are completely 
    separated from the wetting behaviour (green diamonds). 
    For each substrate, the transition to wetting
    is marked by a black solid line
    going through the transitional zone
    between the non-wetting and wetting regions.
    The grey area indicates the presence of air cavity formation,
    with 
    experimental uncertainty indicated by the two dashed lines. 
    It is possible to experimentally determine 
    the maximum Weber number $\We_{\rm m}$ (indicated by the arrow)
    that allows non-wetting behaviour.}
\end{figure}

In the so-called \emph{hovering} behaviour, 
typically observed for droplets of high viscosity liquids, 
an impacting droplet first reaches its 
maximum deformation,
then hovers
and eventually wets the soft substrate. 
In Fig.\:\ref{figure1}d, we show a plot of $h$ versus $t$ 
for this behaviour  
in a representative experiment whereby
$\We$ and $E$ are similar to those in Fig.\:\ref{figure1}c, 
but the Ohnesorge increases from $\Oh = 0.011$
to $0.745$.
In this case, the droplet 
reaches its maximum and retracts 
without either detaching
from the surface or oscillating. 
The hovering state in this case, therefore, 
differs from that of the bouncing-hovering behaviour
in that there is no periodic deformation of the droplet. 
As a result, 
the air film under the droplet is not replenished
and irreversibly drained, leading to 
its rupture at $t = 3.6$. 

\begin{figure*}[t!]
    \centering
    \includegraphics[width=1.8\columnwidth]{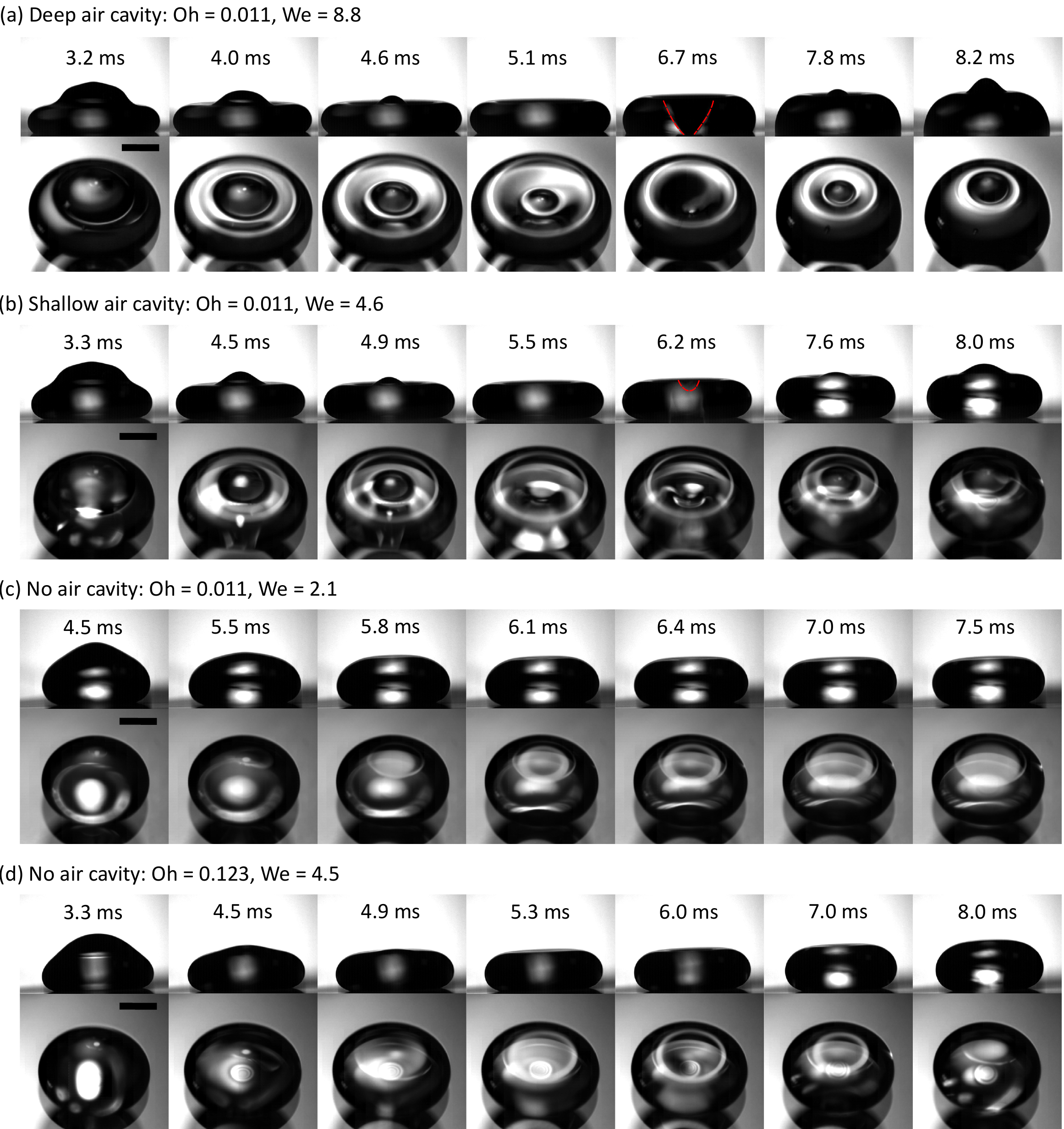}
    \caption{\label{figure3}
     Series of side- and top-view snapshots
     of representative impact experiments showing how air cavity formation 
     is detected. All impact experiments use 
     substrates with $E = 71\,$kPa.
     Experiments in (a-c) use the same liquid ($\Oh=0.011$) and 
     in (d), a more viscous liquid ($\Oh=0.123$) is used.
     The Weber number is progressively reduced from (a) to (c). 
     (a) $\We = 8.8$, a deep air cavity is detected from the side view 
     at 6.7\,ms after the droplet touches the surface.  
     (b) $\We = 4.6$, air cavity formation is observed, 
     although not as deep as the one showed in (a). 
     The red dashed lines in (a) and (b) indicate the contours of the air cavities. 
     (c) $\We = 2.1$, no air cavity is observed from 
     both the side and top views. 
     (d) Similar $\We$ to that in (b), but
     $\Oh$ increases from 0.011 to 0.123,
     both capillary waves and air cavity are mostly suppressed.
     All scale bar represents 1$\,$mm.}
\end{figure*}


We highlight the difference 
between the wetting behaviour (Fig.\:\ref{figure1}b) 
and hovering behaviour (Fig.~\:\ref{figure1}d). 
Although wetting eventually occurs for all impacts 
due to draining of the air film 
separating an impacting droplet and a surface, 
wetting is initiated at widely varying durations after impact.
In the wetting regime, 
wetting occurs within a very short time after impact, e.g., $t = 0.33$ in Fig.\:\ref{figure1}b, 
and is similar to the non-dimensionalised crashing time $t_c = D/(V\tau)$, 
e.g., $t_c \approx 0.34$ for the experiment shown in Fig.\:\ref{figure1}b.
In contrast,
wetting in the hovering regime 
is initiated at $t=3.6$, 
far beyond the crashing time $t_c \approx 1.08$. 
Furthermore, the droplet in the hovering regime 
goes through a duration in which it 
appears floating on the surface 
without any significant oscillation.

\subsection{Transition to wetting of impacting droplets on soft substrates}

We now focus on the conditions enabling the transition to wetting from the 
non-wetting behaviours, which include those exhibiting 
either bouncing-hovering 
or hovering characteristics. 
In Fig.\:\ref{figure2}, 
we show phase diagrams of the 
characteristic behaviours 
obtained on a relatively rigid substrate, with 
$E = 415\,$kPa (Fig.\:\ref{figure2}a) 
and a much softer one, 
with 
$E = 71\,$kPa (Fig.\:\ref{figure2}b).
The phase diagrams are obtained in the $(\Oh,\We)$ parameter space by
varying $\We$ 
between $0.9$ and $11$ 
and $\Oh$ between
$0.011$ and $1.973$.
Typically, we note that the 
non-wetting and wetting regimes are well separated,
and the transition to wetting 
can be experimentally determined at the dividing line between 
these regimes. 
For the impact behaviours obtained on the 
substrate having $E = 415\,$kPa,
the Weber number at the wetting transition 
decreases monotonically with increasing $\Oh$, 
as indicated by the black thick line in Fig.\:\ref{figure2}a. 
To understand this dependence, 
we note that increasing liquid viscosity, or equivalently increasing $\Oh$,
reduces the volume of air trapped under impacting droplets \cite{zhao_impact_2017}
and subsequently facilitates wetting initiation. 
As a result, 
impacting droplets having higher $\Oh$ 
requires lower $\We$
to transition to wetting. 
This is 
also consistent with our experimental data obtained on  
other relatively stiff substrates, i.e., $E\ge265\,$kPa.

For softer substrates, exemplified by the one shown
in Fig.~\ref{figure2}b with $E= 71\,$kPa, 
the transition to wetting generally occurs at a higher $\We$ 
as the soft substrate acts as a cushion 
allowing the trapped air film to sustain higher impact energy 
without being ruptured. 
Interestingly, 
we observe a non-intuitive change 
in the wetting transition for the softer substrate ($E = 71\,$kPa)
compared 
to its rigid counterpart ($E = 415\,$kPa).
At low Ohnesorge number, the required Weber number for wetting transition
first increases with $\Oh$ to reach 
a peak value at $\Oh = 0.262$
and then reduces gradually to reach a plateau.
The change 
in the wetting transition on soft substrates
becomes more pronounced with decreasing $E$.
This suggests another mechanism causing 
wetting initiation other than viscosity-induced 
reduction in volume of the 
trapped air film under impacting droplets. 

We seek to understand
the increase in 
the required $\We$
for wetting transition, i.e., the positive slope portion, by noting 
that there is a region in 
the phase diagram 
in which 
air cavity formation is observed from top surfaces of impacting droplets 
during the spreading phase 
(see Fig.~\ref{figure3}). 
This region overlaps with the
positive slope portion of the wetting transition, 
suggesting a connection between
air cavity formation and wetting initiation. 
We note that this type of air cavities
forms after the top of an impacting droplet
takes 
a pyramidal shape, which 
is caused by convergence of  
capillary waves 
travelling upwards from the 
impact point \cite{renardy_pyramidal_2003,bartolo_singular_2006,mitra_bouncing--wetting_2021,chen_evolution_2011}.
Numerous 
droplet impact studies
reported the occurrence of such air cavities and showed
that they were the enabling factor of 
various phenomena,
ranging from jet formation to dimple inversion \cite{renardy_pyramidal_2003,bartolo_singular_2006,mitra_bouncing--wetting_2021,chen_evolution_2011}.
There are also strong evidences showing 
that air cavity development causes rupture of the air film trapped 
between an impacting droplet and a solid surface 
\cite{pack_failure_2017,mitra_bouncing--wetting_2021}.
This can be connected to a pressure surge at the 
impact point when the droplet reaches its maximum deformation
\cite{renardy_pyramidal_2003,nishimura_dynamics_2018}.
In our experiment, we observe air cavity development
for $\We \geq 4.3$ at the lowest value of 
Ohnesorge number ($\Oh=0.011$), regardless of 
the substrate's elasticity. 
With increasing $\Oh$,
the required $\We$ for air cavity formation 
increases, as indicated by the grey thin lines 
in both Fig.~\ref{figure2}a and b.
Remarkably, the decrease in the dependence of $\We$ on $\Oh$ 
at the wetting transition 
occurs when air cavity formation is absent.

\begin{figure}[t!]
    \centering
    \includegraphics[width=\columnwidth]{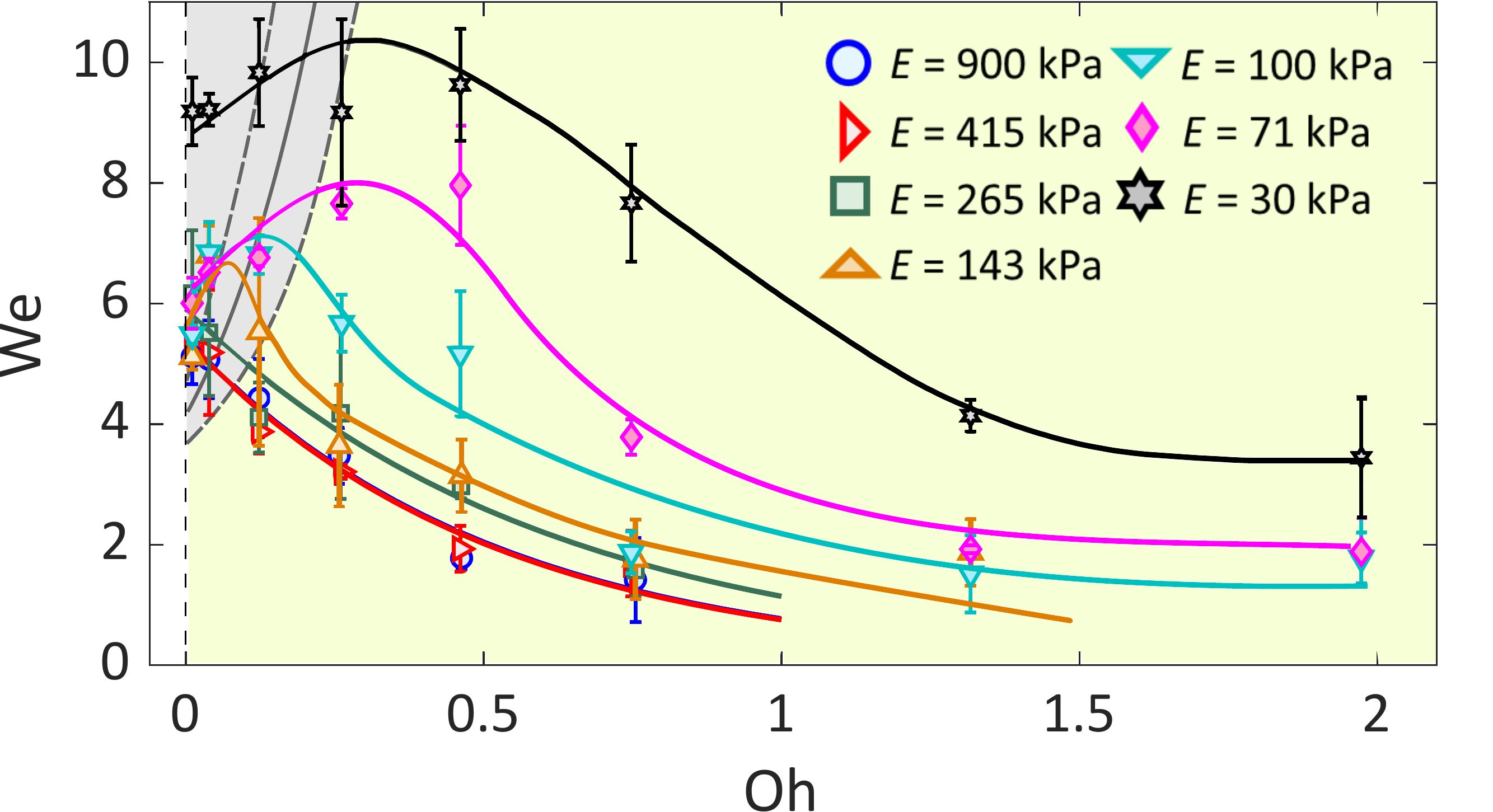}
    \caption{\label{figure4}
    The transition to wetting, plotted as the Weber number
    $\We$
    at the transition versus the Ohnesorge number $\Oh$ 
    for Young's modulus $E$ varying 
    in the range $30\,{\rm kPa} \le E \le 900\,{\rm kPa}$.
    As the elasticity decreases, from $E = 900\,$kPa
    and $E = 415\,$kPa
    to those of softer surfaces,
    the plot shows that the transition to wetting 
    changes both quantitatively (increases in transitional Weber number)
    and qualitatively (from monotonic to non-monotonic behaviours). 
    The grey area indicates the presence of air cavity formation
    for all substrates,
    with 
    experimental uncertainty indicated by the two dashed lines. }
\end{figure}

We attribute the occurrence of air cavity to 
the positive slope portion of $\We$ at the wetting transition. 
We note that air cavity is formed 
when the capillary waves traveling to the top without being 
significantly damped \cite{renardy_pyramidal_2003,vo_droplet_2021,vo_dynamics_2021}, 
thus increasing the liquid viscosity, or equivalently $\Oh$,
requires higher $\We$ to generate and sustain the waves
until they reach the droplet's top apex.
In Fig.~\ref{figure3}b and Fig.~\ref{figure3}d,
we show how capillary waves are diminished by keeping 
the same Weber number ($\We \approx 4.5$)
but increasing $\Oh$ from 0.011 to 0.123. 
Since increasing $\Oh$ diminishes capillary waves and subsequently
air cavity formation,
wetting is less likely to occur,
suggesting that
impacts with higher $\Oh$ requires higher $\We$ 
to transition to the wetting regime. 
Indeed, for all of our substrate's elasticity, 
the positive slope in the dependence of 
$\We$ on $\Oh$ for wetting transition
is mostly associated with air cavity formation, as shown in 
Fig.~\ref{figure4}.
Beyond the range of Ohnesorge number 
that allows capillary waves and air cavity formation, 
the transition to wetting 
recovers its typical characteristics, i.e., monotonically 
decreases in 
the required $\We$ with increasing 
$\Oh$.

To further understand 
how the transition to wetting changes with softer substrates, 
we show 
in the $(\Oh,\We)$ phase diagram
all the transitions to wetting obtained on 
substrates with the Young's modulus varying 
in the range $30\,{\rm kPa} \le E \le 900\,{\rm kPa}$ 
(Fig.\:\ref{figure4}).
In this phase diagram, we also show the 
region in which air cavity formation is observed for all explored values of $E$. 
First, we observe that the required Weber number to transition to wetting 
is lowest for the most rigid substrates, 
e.g., for $E = 900\,$kPa and $E = 415\,$kPa,
the maximum Weber number allowing non-wetting impacts
is $\We_m \approx 5$ at $\Oh=0.011$, similar to 
the Weber number allowing air cavity formation ($\We = 4.3\pm0.4$).
With softer substrates, 
non-wetting impacts are possible at higher Weber number,
which facilitates air cavity formation. 
As a result, softening the substrate affects 
the wetting transition in an indirect 
way.
On one hand, a sufficiently soft substrate acts as a cushion allowing 
the trapped air film 
to sustain higher impact energy without ruptures. 
On the other hand, higher impact energy (equivalently higher $\We$)
facilitates air cavity formation. 
The overlap between wetting transition and air cavity formation
occurs at the limit of low Ohnesorge number and 
causes 
the increase in the required $\We$ for wetting transition.  

\begin{figure}[t]
    \centering
    \includegraphics[width=\columnwidth]{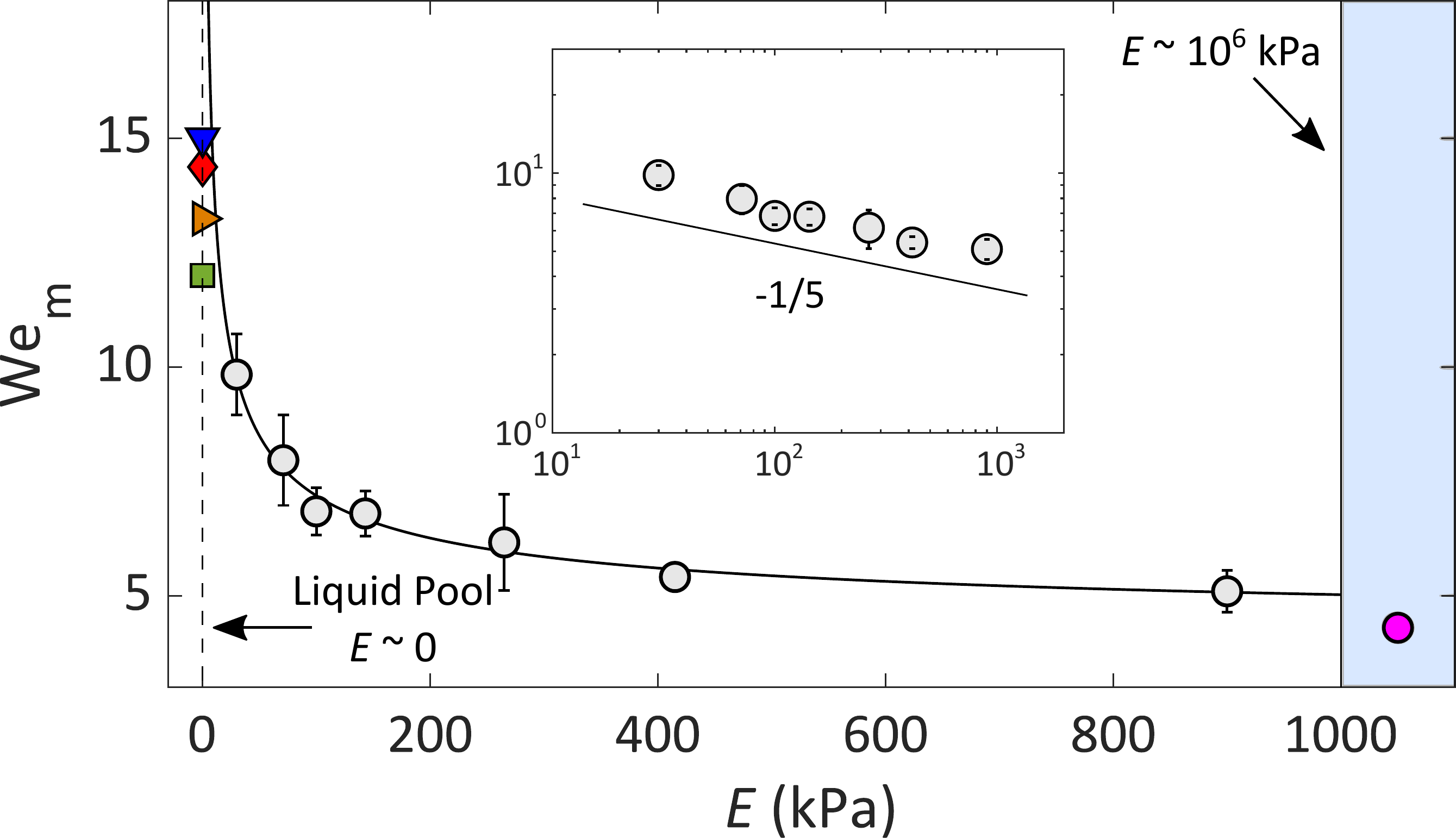}
    \caption{\label{figure5}
    Maximum Weber numbers $\rm{We_{m}}$ allowing non-wetting impacts
    versus Young's modulus $E$. 
    We include data for 
    non-wetting droplet impingement on a 
    liquid pool ($E \sim 0$): 
    1-propanol droplets on 
    1-propanol liquid pool (blue downward triangle)\cite{zhao_transition_2011}, 
    Heptadecane droplet on Heptadecane liquid pool (red diamond) 
    \cite{tang_bouncing_merging_2018}, 
    Tetradecane droplet on Tetradecane liquid pool 
    (yellow right triangle) \cite{tang_bouncing_2019}, 
    and water droplet on water pool (green square) \cite{wu_small_2020}. 
    Non-wetting impact of liquid droplets 
    on a rigid surface ($E \sim 10^6\,$kPa) 
    is marked by a pink circle \cite{de_ruiter_wettability-independent_2015}. 
    The black solid line represent the scaling law $\We_{\rm m} \sim \it{E}^{-1/5}$.
    Inset: our experimental data and the scaling law $\We_{\rm m} \sim \it{E}^{-1/5}$
    in log-log scales.}
\end{figure}

\subsection{Maximum impact energy for non-wetting behaviour}
%
%
%
%
%
In this section, we focus on the maximum impact energy, i.e., measured by the Weber number, 
that non-wetting behaviour is still possible. 
For stiff substrates
where the transition to wetting monotonically decreases with $\Oh$, 
the maximum Weber number allowing non-wetting behaviour
is found at 
the smallest value of $\Oh$, i.e., 
$\Oh = 0.011$.
For relatively softer substrates
where the transition to wetting first increases and then decreases, 
$\We_{\rm m}$ is found 
at the peak of the transition. 
We note that  
for a particular substrate with Young's modulus $E$,
the maximum Weber number 
at the wetting transition
$\We_{\rm m}$ is determined experimentally 
using the $(\Oh,\We)$ phase diagram
(see Fig.~\ref{figure2}). 
As a result, 
$\We_m$ depends on the specific liquid (or $\Oh$) 
and represents 
the substrate's ability to repel droplets without wetting. 

In Fig.\:\ref{figure5}, 
we show how $\rm{We_{m}}$ depends on substrate elasticity 
$E$ and how this dependence compares with data obtained 
on the two extreme cases: rigid surfaces with $E \sim 10^6\,$kPa 
\cite{de_ruiter_wettability-independent_2015}
and liquid layers with $E \sim 0$
\cite{zhao_transition_2011,tang_bouncing_merging_2018,tang_bouncing_2019,wu_small_2020}.
We observe that $\We_{\rm m}$ unequivocally reduces with increasing $E$,
confirming that soft substrates repel liquid droplets 
more efficiently than their rigid counterparts. 
We also find that the reduction of $\We_{\rm m}$ with $E$
is consistent with the scaling law $\We_{\rm m} \sim {\it E}^{-1/5}$, 
as shown in the inset of Fig.~\ref{figure5}. 
Our data obtained for $E$ in the range 
from $30\,$kPa to $900\,$kPa nicely bridge 
the two extreme cases of surface stiffness, from 
the case of ideally negligible elasticity ($E\sim0$) 
and high chance of droplet rebounce
to the one of undeformable surface ($E\sim10^6\,$kPa)
with low chance of rebounce.

\section{Concluding remarks}
In summary, we have experimentally studied 
the impact characteristics of glycerol-water mixture droplets 
on PDMS substrates 
with Young's modulus varying from $30\,$kPa to $900\,$kPa. 
By varying the mixture viscosity 
from 5.0$\,$mPa$\,$s to 
905.7$\,$mPa$\,$s 
(equivalently $\Oh$ 
from 0.011 to 1.973) 
and the impact velocity from 
0.17$\,$ms$^{-1}$ to 0.60\,ms$^{-1}$
(or $\We$ from 0.9 to 16.0),
we found three characteristic behaviours 
for a fixed value of elasticity: wetting, bouncing-hovering, i.e., 
an impacting droplet bounces a few times before 
transitioning to hovering above the surface,
and hovering.
Most importantly, we found that
the Weber number at the transition to wetting decreases 
monotonically with increasing $\Oh$
for substrates at the upper bound of rigidity ($E \ge 265\,\rm{kPa}$). 
For softer substrates ($E \le 143\,\rm{kPa}$), 
the transition to wetting both increases 
in Weber number and changes 
its characteristic to a non-monotonic dependence on $\Oh$:
first it increases with $\Oh$ to reach a peak value, then gradually 
decreases to a plateau. 
Interestingly, we observe 
a strong link between the increase with $\Oh$ 
at the transition to wetting and air cavity formation in the bulk 
of impacting droplets. 
More thorough theoretical and numerical studies
are however required to understand the interaction between air cavity formation and 
the transition to wetting on soft surfaces. 

Despite the complex dependence of the transition to wetting on $\Oh$,
we found that a maximum Weber number $\We_{\rm m}$ allowing non-wetting behaviour
may serve as an excellent indicator of a surface's capacity in rebouncing 
impinging droplets. This is quantified by the 
elasticity dependence of $\We_{\rm m}$, which was
found consistent with the 
scaling law $\We_{\rm m}\sim {\it E}^{-1/5}$
in the explored range of $E$, from $30\,$kPa to $900\,$kPa. 
This experimental scaling law 
further extends past studies of 
droplets impacting liquid pools \cite{zhao_transition_2011,tang_bouncing_merging_2018,tang_bouncing_2019,wu_small_2020}, 
and those on non-deformable surfaces \cite{de_ruiter_wettability-independent_2015}
and may be useful in applications involving 
actuation of viscous droplets on soft materials.


\section*{Author Contributions}
M.L., Q.V. and T.T. conceived the research idea. M.L. performed the experiment and data analysis. M.L., Q.V., S.M. and T.T. discussed the results and wrote the manuscript.

\section*{Conflicts of interest}
There are no conflicts to declare.

\section*{Acknowledgements}
We acknowledge funding from 
the 
Republic of Singapore's Ministry of Education 
(MOE, grant number MOE2018-T2-2-113)
and
the Agency of Science, Technology and Research (A*STAR) 
through the Pharos initiative (grant number 152 37 00102). 
M.L. is supported by Nanyang President's Graduate Scholarship.



\balance



\providecommand*{\mcitethebibliography}{\thebibliography}
\csname @ifundefined\endcsname{endmcitethebibliography}
{\let\endmcitethebibliography\endthebibliography}{}


\bibliographystyle{rsc} 

\end{document}